\documentclass[twocolumn,fp]{jpsj3}

\usepackage{bm}
\usepackage{amsmath}
\usepackage{color,multirow}

\usepackage{graphicx}

\begin{document}

\title{Quasi-classical Theory of Tunneling Spectroscopy in Superconducting 
Topological Insulator}


\author{Shota Takami, Keiji Yada, Ai Yamakage, Masatoshi Sato, and Yukio Tanaka}
\inst{Department of Applied Physics, Nagoya University, Nagoya 464-8603, Japan}
\abst{
We develop a theory of tunneling spectroscopy in superconducting topological insulator (STI),
i.e. superconducting state of a carrier-doped topological insulator.
Based on the quasi-classical approximation,
we obtain an analytical expression of the energy dispersion of
surface Andreev bound states (ABSs) and a compact formula for
tunneling conductance of normal metal/STI junctions.
The obtained compact formula of tunneling conductance makes the analysis
of experimental data easy.
As an application of our theory,
we study tunneling conductance in Cu$_x$Bi$_2$Se$_3$.
We reveal that our theory reproduces the previous results by Yamakage
{\it et al} [Phys. Rev. B 85, 180509(R) (2012)].
We further study magneto-tunneling spectroscopy
in the presence of external magnetic fields,
where the energy of quasiparticles is shifted by the Doppler effect. 
By rotating magnetic fields in the basal plane, 
we can distinguish between two different topological superconducting states
with and without point nodes, in their magneto-tunneling spectra.
}


\maketitle

\section{Introduction}


Topological superconductors have gathered considerable interests in
condensed matter physics.\cite{Schnyder08,wilczek09,qi11,tanaka12,alicea12}. 
They are characterized by the existence of
surface states stemming from a non-trivial topological structure of
the bulk wave functions.\cite{Schnyder08}
These surface states are a special kind of Andreev bound states (ABSs),
which are known as Majorana fermions.\cite{qi11,tanaka12}
It is believed  that
the non-Abelian statistics of Majorana zero modes
open a new possible way to fault tolerant quantum computation.\cite{Nayak}
There are many works on this topic and several proposals for systems where
topological superconductivity  
is expected to be realized.
\cite{Kitaev,fu08,TYBN08,STF09,STF10,fu09,akhmerov09,law09,tanaka09,alicea10,linder10,oreg10,lutchyn10,TMYYS10,STYY11,Nakosai1,Nakosai2,Nakosai3}

One of the most promising candidates of
topological superconductor is Cu doped Bi$_2$Se$_3$
\cite{hor10,sasaki11,kriener11}.  
Since its host undoped material is a topological insulator,
this material is dubbed as superconducting topological 
insulator (STI). 
Theoretically, Fu and Berg classified the possible pair potentials which are 
consistent with the crystal structures.\cite{fu10}
They considered four different irreducible representations
of gap function, based on the two orbital model governing the low energy
excitations. 

According to the Fermi surface criterion for topological
superconductivity \cite{Sato09,fu10,Sato10}, it has been revealed
theoretically that 
ABSs are generated as Majorana fermion 
for odd-parity pairings  
in the $A_{1u}$, $A_{2u}$ and $E_{u}$ irreducible representations.
Among these pairings, $A_{1u}$ and $E_{u}$ pairings have gapless ABSs in the (111) surface. \cite{hao11,hsieh12,yamakage12}
While the pair potential $\Delta_{2}$ in $A_{1u}$ 
does not have nodes on the Fermi surface, 
the pair potential $\Delta_{4}$ in $E_u$ has point nodes 
on the Fermi surface.
For both pairings,
the resulting ABSs have a structural transition
in the energy dispersion.\cite{yamakage12}
In the tunneling conductance between normal metal / insulator / 
Cu$_x$Bi$_2$Se$_3$ junctions, 
the gapped pairing $\Delta_2$ shows a zero bias conductance peak
(ZBCP) for high and intermediate transparent junctions 
and a zero bias conductance dip (ZBCD) for low transparent junctions.
\cite{yamakage12}
On the other hand, $\Delta_{4}$ always shows a ZBCP.
\cite{yamakage12}

In experiments, a pronounced ZBCP has been reported in point contact
measurements of Cu$_{x}$Bi$_{2}$Se$_{3}$.
\cite{sasaki11}
While there are other reports which have observed similar ZBCPs 
\cite{koren11,kirzhner12,koren12,koren13},
there also exist
conflicting results \cite{peng13,Levy}, where
a simple U-shaped
tunneling conductance without ZBCP has been reported 
by the scanning tunneling microscope (STM). 
For the latter experiments, however, a recent theoretical study of
proximity effects on STI  
has suggested that the simple U-shaped spectrum is not explained by an
$s$-wave superconductivity of Cu$_{x}$Bi$_{2}$Se$_{3}$\cite{Mizushima}.
Although there are several studies on 
electronic properties of superconducting states in
Cu$_{x}$Bi$_{2}$Se$_{3}$, 
the pairing symmetry of this material 
has not been clarified yet.
\cite{yamakage13,Hashimoto2013,Yip2013,Nagai1,Nagai2,Balatsky,Zocher,Chen}

Since the experiments of tunneling
spectroscopy have not been  fully settled at present, it is desired to
derive a compact and simple formula of tunneling spectroscopy of STI. 
Indeed, the previous theory of tunneling spectroscopy needs a
complicated numerical calculation,  and thus it is not sufficiently 
convenient to fit experimental data. 
It is not so easy to grasp an intuitive picture of STI as well.
\cite{yamakage12}
To improve them, we use here the quasi-classical theory of STI.
Although there are two orbitals in the microscopic Hamiltonian, the resulting
Fermi surface of STI is rather simple.
Hence, 
it has been proposed to construct a quasi-classical theory of 
STI by extracting low energy degrees of freedom on the Fermi surface.
\cite{Yip2013,Nagai1,Nagai2}

If we can derive a more convenient  theory of tunneling conductance 
by using the quasi-classical approximation, 
our understanding on the tunneling spectroscopy of 
STI can be much more clear since the intuitive 
picture on the relation between ABS\cite{ABS,ABSb,Hu} 
and tunneling conductance 
is expected to be obtained.\cite{TK95,kashiwaya00}
We also expect that the theory is useful like the quasiclassical theory of
charge transport in $p$-wave superconductor junctions 
\cite{YTK97,YTK98,Honerkamp,Proximityp}

\par

In this paper, starting from 
a microscopic Hamiltonian of topological insulators, 
we develop the quasiclassical theory of
tunneling spectroscopy for STI.  
We derive analytical formulas of ABSs and 
tunneling conductance for normal metal / STI junctions. 
Using the obtained formula of ABSs, 
the transition in spectrum of ABS, which was reported in
Ref.\citen{yamakage12},  
is reproduced.  
We also calculate the magneto-tunneling
conductance
in order to   
extract an information on momentum dependence of pair potentials 
from tunneling spectroscopy.\cite{Tanuma2002a,Tanuma2002b,Tanaka2002}
It is found that we can distinguish between 
$\Delta_{2}$ and $\Delta_{4}$,  
although a similar ZBCP appears for both pairings.  
By rotating magnetic fields on the 
basal plane parallel to the interface, $\Delta_{4}$ exhibits 
a two-fold symmetry in the tunneling conductance 
due to the existence 
of point nodes on the Fermi surface.

\section{Model and Formulation}
\subsection{Model Hamiltonian for STI}
To study  carrier-doped topological insulators,
we start from the two-orbital model
proposed to describe Bi$_2$Se$_3$.\cite{S-C_Zhang09,S-C_Zhang10}
The normal-state Hamiltonian is given by
\begin{align}
\hat H_n({\bm k})&= c({\bm k})+m({\bm k})\sigma_x+v_zk_z\sigma_y+v\sigma_z(k_xs_y-k_ys_x),\nonumber\\ \\
m({\bm k}) &= m_0 + m_1 k_z^2 + m_2 k_\parallel^2 ,\\
c({\bm k}) &= -\mu_S + c_1 k_z^2 + c_2 k_\parallel^2 ,
\end{align}
where $k_\parallel=\sqrt{ k_x^2 + k_y^2 }$.
$s_i$ and $\sigma_i$ denote the Pauli matrices in the spin and orbital
spaces, respectively.
In the superconducting state, the BdG Hamiltonian is given by
\begin{eqnarray}
\check H_s({\bm k})&=&
\begin{pmatrix}
\hat H_n({\bm k})&\hat \Delta_\ell\\
\hat \Delta_\ell^\dag&-\hat H_n({\bm k})
\end{pmatrix},
\end{eqnarray}
where $\ell$ labels the type of the pair potential.
In a weak-interaction, where Cooper pairs are formed inside a unit cell,
$\hat \Delta_\ell$ does not have ${\bm k}$-dependence.
In this case, there are six types of pair potentials:
$\hat \Delta_{1a} = \Delta$, $\hat \Delta_{1b} = \Delta \sigma_x$, $\hat \Delta_2 = \Delta \sigma_y s_z$, $\hat \Delta_3 = \Delta \sigma_z$, and 
$\hat \Delta_4 = \{\Delta \sigma_y s_x, \Delta \sigma_y s_y\}$.
$\hat \Delta_{1a}$ and $\hat \Delta_{1b}$ belong to the $A_{1g}$ irreducible representation,
and $\hat \Delta_{2}$, $\hat \Delta_{3}$ and $\hat \Delta_{4}$ belong to the $A_{1u}$, $A_{2u}$ and $E_{u}$ irreducible representations, respectively.
We choose $\Delta \sigma_y s_x$ for $\hat \Delta_{4}$ in this paper.
The results for $\Delta \sigma_y s_y$ is obtained by four-fold rotation around $z$-axis, $(k_x, k_y)\rightarrow(k_y, -k_x)$. \par
Before making a quasiclassical wave function in superconducting state,
we first diagonalize the normal state Hamiltonian.
\begin{eqnarray}
\hat U^\dagger({\bm k}) \hat H_n({\bm k})  \hat U({\bm k})
&=& 
c({\bm k})\tilde s_0\tilde \sigma_0+\eta({\bm k})\tilde s_0\tilde \sigma_z,
\end{eqnarray}
with $\eta({\bm k}) = \sqrt{ m({\bm k})^2 + v_z^2 k_z^2 + v^2 k_\parallel^2 }$.
$\tilde s_0$ labels spin helicity, and $\tilde \sigma_0$ and $\tilde \sigma_z$ represent the band index.
Here, we consider electron-doped Bi$_2$Se$_3$-type topological insulator
where only the conduction band has a Fermi surface.
In addition, the magnitude of the superconducting energy gap is far smaller than the bulk band gap.
Actually, in Cu$_x$Bi$_2$Se$_3$,
the critical temperature ($T_c \simeq 4 \ \rm K$\cite{hor10}) is much smaller 
than the band gap ($0.3 \rm \ eV$\cite{wray}).
In this case, the coherence length $\xi$ is much longer than the inverse of the Fermi wavenumber $ 2 \pi / k_F $, and the quasiclassical approximation is valid.\cite{Eilenberger}
Then, the intraband pairing in valence band and interband pairing between conduction and valence bands can be neglected.
Then, the $8 \times 8$ Bogoliubov de-Gennes Hamiltonian is reduced to $4 \times 4$ one
by extracting only the components of conduction band.
\begin{eqnarray}
\hat H_{\rm eff}({\bm k}) &=& 
\begin{pmatrix}
E_c({\bm k}) {\bm I}
& {\bm \Delta}_\ell({\bm k}) \\
{\bm \Delta}^\dag_\ell({\bm k}) & 
-E_c({\bm k}){\bm I}
\end{pmatrix}.
\end{eqnarray}
Here $E_c({\bm k})=c({\bm k})+\eta({\bm k})$ is the dispersion of the conduction band in the normal state.
${\bm I}$ and ${\bm \Delta}_\ell({\bm k})$ are $2\times2$ matrices
which describe unit matrix and intraband pairing in conduction band, respectively.
The intraband pair potentials for conduction band are written as 
\begin{align}
{\bm \Delta}_{1a}({\bm k})=&
\begin{pmatrix}
\Delta&0\\
0&\Delta
\end{pmatrix},
\\
{\bm \Delta}_{1b}({\bm k})=&
\begin{pmatrix}
\Delta_{1b,0}({\bm k})&0\\
0&\Delta_{1b,0}({\bm k})
\end{pmatrix},
\\ 
{\bm \Delta}_2({\bm k})=&
\begin{pmatrix}
0 & \Delta_{2,x}({\bm k}) - i \Delta_{2,y}({\bm k}) \\
\Delta_{2,x}({\bm k}) + i \Delta_{2,y}({\bm k}) & 0
\end{pmatrix},
\\
{\bm \Delta}_3({\bm k})=&
\begin{pmatrix}
\Delta_{3,z}({\bm k}) & 0 \\
0 & -\Delta_{3,z}({\bm k})
\end{pmatrix},
\\
{\bm \Delta}_4({\bm k})=&
\begin{pmatrix}
\Delta_{4,z}({\bm k}) & \Delta_{4,x}({\bm k}) - i \Delta_{4,y}({\bm k}) \\
\Delta_{4,x}({\bm k}) + i \Delta_{4,y}({\bm k}) & -\Delta_{4,z}({\bm k})
\end{pmatrix},
\end{align}
where
\begin{align}
\Delta_{1b,0}=& 
\Delta \frac{m({\bm k})}{ \sqrt{ m({\bm k})^2 + v_z^2 k_z^2 + v^2 k_\parallel^2 } },\label{eq:1b0}\\
\Delta_{2,x}=& 
\Delta \frac{v_z k_z}{ \sqrt{ m({\bm k})^2 + v_z^2 k_z^2 } },\\
\Delta_{2,y}=&
\Delta \frac{m({\bm k}) v k_\parallel}{ \sqrt{ ( m({\bm k})^2 + v_z^2 k_z^2 )(  m({\bm k})^2 + v_z^2 k_z^2 + v^2 k_\parallel^2 ) } },\\
\Delta_{3,z}=& 
\Delta \frac{v k_\parallel}{ \sqrt{ m({\bm k})^2 + v_z^2 k_z^2 + v^2 k_\parallel^2 } },\\
\Delta_{4,x}=&  
\Delta \frac{ m({\bm k}) v k_x }{  \sqrt{ ( m({\bm k})^2 + v_z^2 k_z^2 )(  m({\bm k})^2 + v_z^2 k_z^2 + v^2 k_\parallel^2 ) } },\\
\Delta_{4,y}=& 
-\Delta \frac{v_z k_x k_z }{ k_\parallel \sqrt{ m({\bm k})^2 + v_z^2 k_z^2} },\\
\Delta_{4,z}=& 
-\Delta \frac{ v_z k_y k_z }{ k_\parallel \sqrt{ m({\bm k})^2 + v_z^2 k_z^2 + v^2 k_\parallel^2 } }.
\end{align}
\subsection{Analytical Formula of the Andreev Bound States}
Solving the effective BdG equation
\begin{equation}
\hat{H}_{\rm eff} \Psi_S^\ell (z) = E \Psi_S^\ell (z),
\end{equation}
the wave function for each pair potential is derived as
\begin{align}
\Psi_S^{1a}(z)=&{\bm S} \left\{ c_1 
\begin{pmatrix}
1 \\ 0 \\ \Gamma_{1a} \\ 0 
\end{pmatrix}
e^{iq_z z} + c_2
\begin{pmatrix}
0 \\ 1 \\ 0 \\ \Gamma_{1a}
\end{pmatrix}
e^{iq_z z} \right. \nonumber\\
&
\left. + c_3
\begin{pmatrix}
\Gamma_{1a} \\ 0 \\ 1 \\ 0
\end{pmatrix}
e^{-iq_z z} + c_4
\begin{pmatrix}
0 \\ \Gamma_{1a} \\ 0 \\ 1
\end{pmatrix}
e^{-iq_z z}
\right\}, 
\\
\Psi_S^{1b}(z)=&{\bm S} \left\{ c_1 
\begin{pmatrix}
1 \\ 0 \\ \Gamma_{1b} \\ 0
\end{pmatrix}
e^{i q_z z} + c_2
\begin{pmatrix}
0 \\ 1 \\ 0 \\ \Gamma_{1b}
\end{pmatrix}
e^{i q_z z} \right.\nonumber \\
& 
\left. + c_3
\begin{pmatrix}
\Gamma_{1b} \\ 0 \\ 1 \\ 0
\end{pmatrix}
e^{-i q_z z} + c_4
\begin{pmatrix}
0 \\ \Gamma_{1b} \\ 0 \\ 1 
\end{pmatrix}
e^{-i q_z z}
\right\},
\\
\Psi_S^2(z)=& {\bm S} \left\{ c_1 
\begin{pmatrix}
1 \\ 0 \\ 0 \\ \Gamma_{2+} 
\end{pmatrix}
e^{iq_z z} + c_2 
\begin{pmatrix}
0 \\ 1 \\ \Gamma_{2-} \\ 0
\end{pmatrix}
e^{iq_z z} \right. \nonumber \\
&
 \left. + c_3
\begin{pmatrix}
0 \\ -\Gamma_{2-} \\ 1 \\ 0
\end{pmatrix}
e^{-iq_z z} + c_4
\begin{pmatrix}
-\Gamma_{2+} \\ 0 \\ 0 \\ 1
\end{pmatrix}
e^{-iq_z z}
\right\}, 
 \\
\Psi_S^3 (z) =& {\bm S} \left\{ c_1
\begin{pmatrix}
1 \\ 0 \\ \Gamma_3 \\ 0
\end{pmatrix}
e^{iq_z z} + c_2
\begin{pmatrix}
0 \\ 1 \\ 0 \\ -\Gamma_3
\end{pmatrix}
e^{iq_z z} \right. \nonumber \\
&
 \left. + c_3
\begin{pmatrix}
\Gamma_3 \\ 0 \\ 1 \\ 0
\end{pmatrix}
e^{-iq_z z} + c_4
\begin{pmatrix}
0 \\ -\Gamma_3 \\ 0 \\ 1
\end{pmatrix}
e^{-iq_z z}
\right\},
 \\
\Psi_S^4 (z)=& {\bm S} \left\{ c_1
\begin{pmatrix}
1 \\ 0 \\ \Gamma_4 \\ \Gamma_{4+}
\end{pmatrix}
e^{iq_z z} + c_2 
\begin{pmatrix}
0 \\ 1 \\ \Gamma_{4-} \\ -\Gamma_4
\end{pmatrix}
e^{iq_z z} \right. \nonumber  \\
&
\left. + c_3
\begin{pmatrix}
-\Gamma_4 \\ \Gamma_{4-} \\ 1 \\ 0
\end{pmatrix}
e^{-iq_z z} + c_4 
\begin{pmatrix}
\Gamma_{4+} \\ \Gamma_4 \\ 0 \\ 1
\end{pmatrix}
e^{-iq_z z}
\right\}.
\end{align}
where  $q_z > 0$ is the Fermi momentum defined by $E_c(q_z) = 0$.
Here the matrix
\begin{eqnarray}
{\bm S} = \frac{1}{\sqrt{2}}
\begin{pmatrix}
1 & 1 & 0 & 0 \\
e^{i\phi_k} & -e^{i\phi_k} & 0 & 0 \\
0 & 0 & 1 & 1 \\
0 & 0 & e^{i\phi_k} & -e^{i\phi_k}
\end{pmatrix},\\
\cos \phi_k = \frac{ k_x }{ k_\parallel } ,\ \sin \phi_k = - \frac{ k_y }{ k_\parallel },
\end{eqnarray}
is attached to restore the transformation of the spin basis. 
\begin{eqnarray}
\Gamma_{1a} &=& 
\left\{
\begin{array}{ll}
\frac{ \Delta }{  E + \sqrt{E^2 - \Delta^2} } & ( \ \Delta < E\ ) \\
\frac{ \Delta }{  E + i \sqrt{ \Delta^2 - E^2} } & (\ |E| < \Delta\ ) \\
\frac{ \Delta }{  E - \sqrt{E^2 - \Delta^2} } & ( \ E < -\Delta \ ) \\
\end{array}
\right. \\
\Gamma_{1b} &=& 
\left\{
\begin{array}{ll}
\frac{ \Delta_{1b,0} }{  E + \sqrt{E^2 - \Delta_{1b,0}^2} } & ( \ |\Delta_{1b,0}| < E\ ) \\
\frac{ \Delta_{1b,0} }{  E + i \sqrt{ \Delta_{1b,0}^2 - E^2} } & (\ |E| < |\Delta_{1b,0}|\ ) \\
\frac{ \Delta_{1b,0} }{  E - \sqrt{E^2 - \Delta_{1b,0}^2} } & ( \ E < -|\Delta_{1b,0}| \ ) \\
\end{array}
\right. \\
\Gamma_{2\pm} &=& 
\left\{
\begin{array}{ll}
\frac{ \Delta_{2,x} \pm i \Delta_{2,y} }{  E + \sqrt{E^2 - ( \Delta_{2,x}^2 + \Delta_{2,y}^2 ) } } \\
 \hspace{5em} ( \ \sqrt{ \Delta_{2,x}^2 + \Delta_{2,y}^2 } < E\ ) \\
\frac{ \Delta_{2,x} \pm i \Delta_{2,y} }{  E + i \sqrt{ ( \Delta_{2,x}^2 + \Delta_{2,y}^2 ) - E^2} } \\
 \hspace{5em} (\ |E| < \sqrt{ \Delta_{2,x}^2 + \Delta_{2,y}^2 }\ ) \\
\frac{ \Delta_{2,x} \pm i \Delta_{2,y} }{  E - \sqrt{E^2 - ( \Delta_{2,x}^2 + \Delta_{2,y}^2 ) } } \\
 \hspace{5em} ( \ E < -\sqrt{ \Delta_{2,x}^2 + \Delta_{2,y}^2 } \ ) \\
\end{array}
\right. \\
\Gamma_{3} &=& 
\left\{
\begin{array}{ll}
\frac{ \Delta_{3,z} }{  E + \sqrt{E^2 - \Delta_{3,z}^2} } & ( \ |\Delta_{3,z}| < E\ ) \\
\frac{ \Delta_{3,z} }{  E + i \sqrt{ \Delta_{3,z}^2 - E^2} } & (\ |E| < |\Delta_{3,z}| \ ) \\
\frac{ \Delta_{3,z} }{  E - \sqrt{E^2 - \Delta_{3,z}^2} } & ( \ E < -|\Delta_{3,z}| \ ) \\
\end{array}
\right. \\
\Gamma_4 &=& 
\left\{
\begin{array}{ll}
\frac{ \Delta_{4,z} }{  E + \sqrt{E^2 - ( \Delta_{4,x}^2 + \Delta_{4,y}^2 + \Delta_{4,z}^2)} } \\
 \hspace{3em}  ( \ \sqrt{ \Delta_{4,x}^2 + \Delta_{4,y}^2 + \Delta_{4,z}^2 } < E\ ) \\
\frac{ \Delta_{4,z} }{  E + i \sqrt{( \Delta_{4,x}^2 + \Delta_{4,y}^2 + \Delta_{4,z}^2) - E^2} } \\
 \hspace{3em}  (\ |E| < \sqrt{ \Delta_{4,x}^2 + \Delta_{4,y}^2 + \Delta_{4,z}^2 }\ ) \\
\frac{ \Delta_{4,z} }{  E - \sqrt{E^2 - ( \Delta_{4,x}^2 + \Delta_{4,y}^2 + \Delta_{4,z}^2)} } \\
 \hspace{3em}  ( \ E < -\sqrt{ \Delta_{4,x}^2 + \Delta_{4,y}^2 + \Delta_{4,z}^2 } \ ) \\
\end{array}
\right. \\
\Gamma_{4\pm} &=& 
\left\{
\begin{array}{ll}
\frac{ \Delta_{4,x} \pm i \Delta_{4,y} }{  E + \sqrt{E^2 - ( \Delta_{4,x}^2 + \Delta_{4,y}^2 + \Delta_{4,z}^2)} } \\
 \hspace{3em}  ( \ \sqrt{ \Delta_{4,x}^2 + \Delta_{4,y}^2 + \Delta_{4,z}^2 } < E\ ) \\
\frac{  \Delta_{4,x} \pm i \Delta_{4,y} }{  E + i \sqrt{( \Delta_{4,x}^2 + \Delta_{4,y}^2 + \Delta_{4,z}^2) - E^2} } \\
 \hspace{3em}  (\ |E| < \sqrt{ \Delta_{4,x}^2 + \Delta_{4,y}^2 + \Delta_{4,z}^2 }\ ) \\
\frac{  \Delta_{4,x} \pm i \Delta_{4,y} }{  E - \sqrt{E^2 - ( \Delta_{4,x}^2 + \Delta_{4,y}^2 + \Delta_{4,z}^2)} } \\
 \hspace{3em}  ( \ E < -\sqrt{ \Delta_{4,x}^2 + \Delta_{4,y}^2 + \Delta_{4,z}^2 } \ ) \\
\end{array}
\right. 
\end{eqnarray}
We calculate the energy spectrum of the ABS from these wave functions
for semi-infinite Cu$_x$Bi$_2$Se$_3$ with $ z > 0 $ by imposing the boundary condition $\Psi_S^\ell (0) = 0$.
It is clarified that there is no ABSs in $ \Delta_{1a},\ \Delta_{1b}$ and $\ \Delta_3$.
On the other hand,
there exists the ABS in $ \Delta_2$ and $ \Delta_4 $ whose energy spectra are expressed as
\begin{align}
E_b(k_x, k_y)
=& 
\pm \Delta  \frac{ v k_\parallel }{ \sqrt{ m({\bm k'})^2 + v_z^2 q_z^2 + v^2 k_\parallel^2 } } \frac{ m({\bm k'}) }{ \sqrt{ m({\bm k'})^2 + v_z^2 q_z^2  } } ,\label{eq32}\\
E_b(k_x, k_y)
=&
\pm \Delta \frac{ v k_x }{ \sqrt{ m({\bm k'})^2 + v_z^2 q_z^2 + v^2 k_\parallel^2 } } \frac{ m({\bm k'}) }{ \sqrt{ m({\bm k'})^2 + v_z^2 q_z^2 } }\label{eq33},
\end{align}
with ${\bm k'} = (k_x, k_y, q_z)$.
Equations (\ref{eq32}) and (\ref{eq33}) are one of the main results of 
the present paper.
In the later section,
we explain that the unconventional caldera-type or Ridge-type 
dispersion of the ABS are produced by $m({\bm k})$.\cite{Yip2013}

\subsection{Analytical Formula of the Conductance}
Next, we calculate tunneling conductance between Cu$_x$Bi$_2$Se$_3$ ($z>0$) and normal metal ($ z<0 $).
For normal metal, we consider a single band model with parabolic dispersion
$E_n({\bm k})=\hbar^2k^2/2m_N-\mu_N$.
In the normal metal, the wave function is written as
\begin{eqnarray}
\Psi_N^1 ( z ) &=&
\begin{pmatrix}
1 \\ 0 \\ 0 \\ 0
\end{pmatrix}
e^{ik_z z} + a_1 
\begin{pmatrix}
0 \\ 0 \\ 1 \\ 0
\end{pmatrix}
e^{ik_z z} + \overline{a}_1
\begin{pmatrix}
0 \\ 0 \\ 0 \\ 1 
\end{pmatrix}
e^{ik_z z}\nonumber \\
&&
+ b_1
\begin{pmatrix}
1 \\ 0 \\ 0 \\ 0
\end{pmatrix}
e^{-ik_z z} + \overline{b}_1
\begin{pmatrix}
0 \\ 1 \\ 0 \\ 0
\end{pmatrix}
e^{-ik_z z},
\\
\Psi_N^2 ( z )&=&
\begin{pmatrix}
0 \\ 1 \\ 0 \\ 0
\end{pmatrix}
e^{ik_z z} + a_2 
\begin{pmatrix}
0 \\ 0 \\ 1 \\ 0
\end{pmatrix}
e^{ik_z z} + \overline{a}_2
\begin{pmatrix}
0 \\ 0 \\ 0 \\ 1 
\end{pmatrix}
e^{ik_z z}\nonumber \\
&&
+ b_2
\begin{pmatrix}
1 \\ 0 \\ 0 \\ 0
\end{pmatrix}
e^{-ik_z z} + \overline{b}_2
\begin{pmatrix}
0 \\ 1 \\ 0 \\ 0
\end{pmatrix}
e^{-ik_z z},
\end{eqnarray}
for the injection of the spin up and spin down electron.
The four transmission and reflection coefficients are determined by the boundary condition.
By considering the delta function barrier potential $ V(z) = \frac{Z}{2} \delta(z) $, the boundary condition is summarized in the form 
\begin{align}
{\bm b} - {\bm S}_0 {\bm c} + {\bm S}_0 {\bm \Gamma}_- {\bm d}
=& 
{\bm u},\\
{\bm a} + {\bm S}_0 {\bm \Gamma}_+ {\bm c} - {\bm S}_0 {\bm d}
=&
{\bm 0},\\
(-v_N+iZ){\bm b}-v_S{\bm S}_0 {\bm c} - v_S {\bm S}_0 {\bm \Gamma}_- {\bm d}
=&
(v_N+iZ){\bm u},\\
( v_N + iZ ) {\bm a} + v_S {\bm S}_0 {\bm \Gamma}_+ {\bm c} + v_S {\bm S}_0 {\bm d}
=&
\bm 0,
\end{align}
where $v_N$ and $v_S$ are the Fermi velocities in the $k_z$-direction inside the normal metal and STI, respectively.
They are given by
\begin{align}
v_N &= \frac{\hbar}{m_N}\sqrt{k_{FN}^2-k_\parallel^2}, \\
v_S &= 2c_1k_{Fz}+\frac{(2m({\bm k})m_1+v_z^2)k_{Fz}}{\eta({\bm k})},
\end{align}
where $k_{Fz}$ is defined by the equation
$c({\bm k})+\eta({\bm k})=0$.
The other parameters are defined as
\begin{eqnarray}
{\bm S_0} &\equiv& \frac{1}{\sqrt{2}}
\begin{pmatrix}
1 & 1 \\
e^{i\phi_k} & -e^{i\phi_k}
\end{pmatrix},
\end{eqnarray}
\begin{eqnarray}
{\bm \Gamma_\pm} &\equiv&
\left\{
\begin{array}{ll}
\begin{pmatrix} -\Gamma_{1a} & 0 \\ 0 & -\Gamma_{1a} \end{pmatrix} & \mbox{for }\Delta_{1a} \\ \\
\begin{pmatrix} -\Gamma_{1b} & 0 \\ 0 & -\Gamma_{1b} \end{pmatrix} & \mbox{for }\Delta_{1b} \\ \\
\begin{pmatrix} 0 & \mp\Gamma_{2\mp} \\ \mp\Gamma_{2\pm} & 0 \end{pmatrix} & \mbox{for }\Delta_2 \\ \\
\begin{pmatrix} -\Gamma_3 & 0 \\ 0 & \Gamma_3 \end{pmatrix} & \mbox{for }\Delta_3 \\ \\
\begin{pmatrix} \mp\Gamma_4 & -\Gamma_{4\mp} \\ -\Gamma_{4\pm} & \pm\Gamma_4 \end{pmatrix} & \mbox{for }\Delta_4
\end{array}
\right. 
\end{eqnarray}
\begin{eqnarray}{\bm a} &\equiv&
\begin{pmatrix}
a_1 \\ \overline{a}_1
\end{pmatrix}
\mbox{for }\Psi_N^1(z)\ \ \ \  {\rm or}\ 
\begin{pmatrix}
a_2 \\ \overline{a}_2
\end{pmatrix}
\mbox{for }\Psi_N^2(z),
\\
{\bm b} &\equiv&
\begin{pmatrix}
b_1 \\ \overline{b}_1
\end{pmatrix}
\mbox{for }\Psi_N^1(z)\ \ \ \  {\rm or}\ 
\begin{pmatrix}
b_2 \\ \overline{b}_2
\end{pmatrix}
\mbox{for }\Psi_N^2(z),
\\
{\bm c} &\equiv&
\begin{pmatrix}
c_1 \\ c_2
\end{pmatrix},
\\
{\bm d} &\equiv&
\begin{pmatrix}
c_3 \\ c_4
\end{pmatrix},
\\
{\bm u} &\equiv&
\begin{pmatrix}
-1 \\ 0
\end{pmatrix}
\mbox{for }\Psi_N^1(z)\ \ \ \  {\rm or}\ 
\begin{pmatrix}
0 \\ -1
\end{pmatrix}
\mbox{for }\Psi_N^2(z).
\end{eqnarray}
By eliminating  ${\bm c}$ and ${\bm d}$ in the boundary condition, then
\begin{eqnarray}
\bm a &=& ( 1 - R ) {\bm S}_0 {\bm \Gamma}_+ ( {\bm I} - R {\bm \Gamma}_- {\bm \Gamma}_+ )^{-1} {\bm S}_0^{-1} {\bm u},\\
\bm b &=& \gamma {\bm S_0}( {\bm I} - {\bm \Gamma}_- {\bm \Gamma}_+ )( {\bm 1} - R {\bm \Gamma}_+ {\bm \Gamma}_- )^{-1} {\bm S}_0^{-1} {\bm u},
\end{eqnarray}
where $\gamma = \frac{ v_S - v_N - iZ }{ v_S + v_N - iZ }$ and $ R = |\gamma|^2 $.
By using these equations, we can aquire the expression of the transmissivity as
\begin{eqnarray}
T_S(\theta, \phi)
&=& 
1 + |{\bm a}|^2 - |{\bm b}|^2 \nonumber \\ 
&=&
\frac{T_N}{2} \left\{ {\bm u}^\dagger {\bm S}_0
( {\bm I} - R {\bm \Gamma}_+^\dagger {\bm \Gamma}_-^\dagger )^{-1} \right. \nonumber \\
&&
\times ( {\bm 1} + T_N {\bm \Gamma}_+^\dagger {\bm \Gamma}_+ - R {\bm \Gamma}_+^\dagger {\bm \Gamma}_-^\dagger {\bm \Gamma}_- {\bm \Gamma}_+ ) \nonumber \\
&&
\times \left. ( {\bm 1} - R{\bm \Gamma}_- {\bm \Gamma}_+ )^{-1} {\bm S}_0^\dagger {\bm u} \right\}.
\end{eqnarray}
Here $T_N = 1-R$ is the transmissivity when $ \rm Cu_x Bi_2 Se_3 $ is in the normal state. If we consider the injection of the spin up electron, $ ^T{\bm u} =  ( -1, 0 )  $, the transmissivity is expressed as
\begin{eqnarray}
T_S(\theta, \phi)
&=&
\frac{T_N}{2} \left\{ {\bm S_0} ( {\bm I} - R {\bm \Gamma}_+^\dagger {\bm \Gamma}_-^\dagger )^{-1} \right. \nonumber \\
&&
\times ( {\bm 1} + T_N {\bm \Gamma}_+^\dagger {\bm \Gamma}_+ - R {\bm \Gamma}_+^\dagger {\bm \Gamma}_-^\dagger {\bm \Gamma}_- {\bm \Gamma}_+ ) \nonumber \\
&&
\left. \times ( {\bm 1} - R{\bm \Gamma}_- {\bm \Gamma}_+ )^{-1} \bm S_0^\dagger  \right\}_{11}.
\end{eqnarray}
On the other hand, if we consider the injection of the spin down electron, $^T{\bm u} =  (0, -1) $, the transmissivity is expressed as
\begin{eqnarray}
T_S(\theta, \phi)
&=&
\frac{T_N}{2} \left\{ {\bm S_0} ( {\bm 1} - R {\bm \Gamma}_+^\dagger {\bm \Gamma}_-^\dagger )^{-1} \right.\nonumber \\
&&
\times ( {\bm 1} + T_N {\bm \Gamma}_+^\dagger {\bm \Gamma}_+ - R {\bm \Gamma}_+^\dagger {\bm \Gamma}_-^\dagger {\bm \Gamma}_- {\bm \Gamma}_+ )\nonumber \\
&&
\times \left. ( {\bm 1} - R{\bm \Gamma}_- {\bm \Gamma}_+ )^{-1} \bm S_0^\dagger  \right\}_{22}.
\end{eqnarray}
As a result, the transmissivity is expressed as
\begin{eqnarray}
T_S(\theta, \phi) 
&=& 
\frac{T_N}{4} {\rm Tr} \left\{ ( {\bm 1} - R {\bm \Gamma}_+^\dagger {\bm \Gamma}_-^\dagger )^{-1} \right. \nonumber \\
&&
\times ( {\bm 1} + T_N {\bm \Gamma}_+^\dagger {\bm \Gamma}_+ - R {\bm \Gamma}_+^\dagger {\bm \Gamma}_-^\dagger {\bm \Gamma}_- {\bm \Gamma}_+ ) \nonumber \\
&&
\times \left. ( {\bm 1} - R{\bm \Gamma}_- {\bm \Gamma}_+ )^{-1}  \right\}.\label{eq53}
\end{eqnarray}
Equation. (\ref{eq53}) is also the main result of the 
present paper. This conductance formula is
a natural extension of 
that for single band unconventional superconductors.\cite{TK95,kashiwaya00}
To derive this expression, we used $ {\bm S_0}^{-1} = {\bm S_0}^\dagger $ and ${\rm Tr} \{ {\bm A}{\bm  B}{\bm C} \} = {\rm Tr} \{ {\bm C}{\bm A}{\bm B} \} $.
This formula of the transmissivity can be simplified as
\begin{eqnarray}
&&T_S(\theta, \phi)=
T_N \frac{ 1 + T_N |\Gamma_{1a}|^2 - ( 1 - T_N ) |\Gamma_{1a}|^4 }{ | 1 - ( 1 - T_N )\Gamma_{1a}^2 |^2 },\\
&&T_S(\theta, \phi)=
T_N \frac{ 1 + T_N |\Gamma_{1b}|^2 - ( 1 - T_N ) |\Gamma_{1b}|^4 }{ | 1 - ( 1 - T_N )\Gamma_{1b}^2 |^2 } ,\\
&&T_S(\theta, \phi)=
\frac{T_N}{2} \left\{ \frac{ 1 + T_N |\Gamma_{2+}|^2 - ( 1 - T_N ) |\Gamma_{2+}|^4 }{ | 1 + ( 1 - T_N ) \Gamma_{2+}^2 |^2 } \right.\\
&&
\left. + \frac{ 1 + T_N |\Gamma_{2-}|^2 - ( 1 - T_N ) |\Gamma_{2-}|^4 }{ | 1 + ( 1 - T_N ) \Gamma_{2-}^2 |^2 } \right\},\\
&&T_S(\theta, \phi)=
T_N \frac{ 1 + T_N |\Gamma_3|^2 - ( 1 - T_N ) |\Gamma_3|^4 }{ | 1 - ( 1 - T_N )\Gamma_3^2 |^2 },
\end{eqnarray}
for $\Delta_{1a}$, $\Delta_{1b}$, $\Delta_{2}$ and $\Delta_{3}$, respectively.
It is remarkable that these equations are essentially the same as the
standard formula of transmissivity \cite{TK95}. 
Using this transmissivity, the normalized tunneling conductance can be calculated by integrating the angle of the injection of the electron
\begin{eqnarray}
\frac{ G_S }{ G_N } = \frac{ \int_0^{2\pi} d\phi \int_0^{\pi/2} d\theta \ \sin 2\theta \ T_S }{ \int_0^{2\pi} d\phi \int_0^{\pi/2} d\theta \ \sin 2\theta  \ T_N }.
\end{eqnarray}

In the following,  we also consider Doppler effect in the presence of magnetic fields parallel to the $xy$-plane. 
If we denote the angle $\gamma$ between the magnetic field and 
the $x$-axis, the magnetic field in the superconductor 
is given by
\begin{eqnarray}
{\bm H}( {\bm r} ) = ( H e^{-z/\lambda} \cos \gamma, H e^{-z/\lambda} \sin \gamma, 0 )
\end{eqnarray}
for $z>0$ with penetration length 
$\lambda$.  
The vector potential is given by
\begin{eqnarray}
{\bm A} ( {\bm r} ) = ( H \lambda e^{-z/\lambda} \sin \gamma, -H \lambda e^{-z/\lambda} \cos \gamma, 0 )
\end{eqnarray}
In order to calculate the tunneling conductance, 
it is sufficient to know the value only near the 
interface with $ z \ll \lambda $. 
Then, the vector potential can be approximated as 

\begin{eqnarray}
{\bm A} ( {\bm r} ) \sim ( H \lambda \sin \gamma, -H \lambda \cos \gamma, 0 )
\end{eqnarray}
The energy of the quasi-particle $E(\bm k)$ is shifted as 
$E({\bm k} - e {\bm A}) 
=E({\bm k}) - e v_\parallel H \lambda \cos(\phi - \gamma)$
by the Doppler effect. 
Here, $v_\parallel$ is the magnitude of the in-plane 
group velocity of quasi-particle and 
$\phi$ is measured from the $x$-axis.

\section{Calculated Results}
In this section, we show the calculated results of the ABSs, conductance and magneto-tunneling conductance.
For the material parameters $v_z$, $v$, $m_0$, $m_2$, $c_1$, $c_2$, we adopt the values for Bi$_2$Se$_3$. \cite{S-C_Zhang10}
Since these parameters proposed in Ref. \citen{S-C_Zhang10} give a cylindrical Fermi surface in the tight-binding model, different parameters are proposed in Ref. \citen{sasaki11} and \citen{Hashimoto2013}.
However, the obtained Fermi surface in the present continuum model is the ellipsoidal one in both cases since the Brillouin zone does not exist.
Though the Fermi momentum and the Fermi velocity along the $z$-direction are different between these two kinds of parameters,
we have confirmed that this difference does not influence the results qualitatively.
\subsection{Andreev Bound State}
\begin{figure}[htbp]
 \begin{center}
  \includegraphics[width = 85 mm]{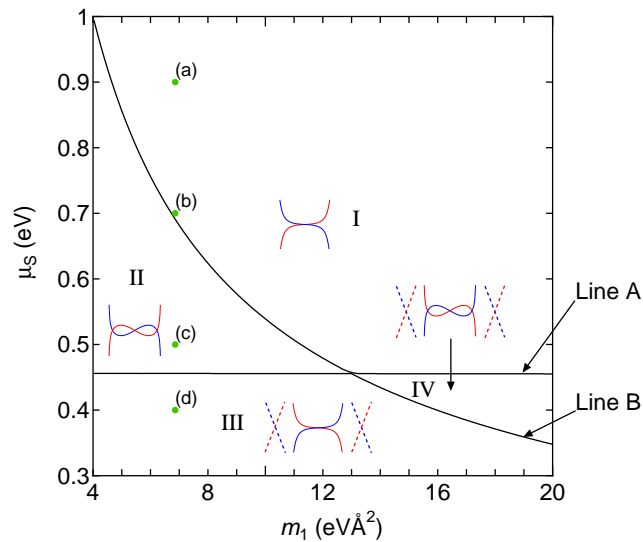}
 \end{center}
 \caption{(color online) Structural change of ABS originating from superconductivity
and surface Dirac state stemming from topological insulator in terms of $\mu_S$ and $m_1$ for $\Delta_2$.
The dotted lines show the surface Dirac state and the solid lines show the ABS. Only the ABS indicated by the solid lines is described in the present quasiclassical approximation.
Line A is the boundary of the structural change of ABS such that the Dirac cone emerges.
Line B is the boundary such that the group velocity at $k_\parallel = 0$ becomes 0.
The parameters we have used in the actual calculations are shown by the points (a)-(d).}
 \label{fig1}
\end{figure}

We first outline important features of ABSs we discuss in this paper.
Figure \ref{fig1} illustrates schematic shapes of the dispersion of the ABS in $\Delta_2$
for various values of $\mu_S$ and  $m_{1}$.
In the region III and IV below line A,
it is known that the surface Dirac cone stemming from topological insulator (dotted lines in Fig. \ref{fig1}) and the 
ABS (solid lines in Fig. \ref{fig1}) are well separated,
and the surface Dirac cone exists outside the Fermi momentum $k_F$. \cite{hao11,yamakage12}
This surface Dirac cone can not be described in the quasiclassical approximation.
On the other hand, in the region I and II,
the ABS merges with the surface Dirac cone. 
The group velocity of ABS at $k_\parallel = 0$ becomes zero on line B. 
In the region I, the shape of the dispersion of the ABS is essentially the same as the standard 
Majorana cone realized in Balian-Werthamer (BW) phase of superfluid $^{3}$He.
In the region II, the group velocity of ABS at $k_\parallel = 0$ is negative and
the shape of the dispersion becomes caldera-type for $\Delta_2$.
\cite{yamakage12}
In this region, the dispersion of the ABS is twisted and it crosses zero energy at finite $k_x$ and $k_y$ as well as at $(k_x, k_y)=(0, 0)$,
since $m({\bm k}')$ can be zero in Eq. (\ref{eq32}).
On the other hand, in the region III, the ABS looks like a conventional Majorana cone again,
because the solution of $m({\bm k}')=0$ moves outside the Fermi energy.
It is noted that the shape of the ABS in the region III is the same as that in the region I,
however, the sign of the group velocity at $k_\parallel=0$ in these two regions are opposite.
In this case,
the surface Dirac state can not be described in the quasi-classical approximation,
but this hardly affects the conductance
because the group velocity of this surface Dirac cone is much larger than that of the ABS.

In the case of $\Delta_4$,
as seen from Eqs. (\ref{eq32}) and (\ref{eq33}),
the dispersion of the ABS along $k_x$-axis is identical with $\Delta_2$.
On the other hand, the energy dispersion of the ABS along $k_y$-axis is zero-energy flat band.
Thus, a ridge-type (valley-type) ABS appears in the region II (region I).

\begin{figure}[htbp]
 \begin{center}
  \includegraphics[width = 70 mm]{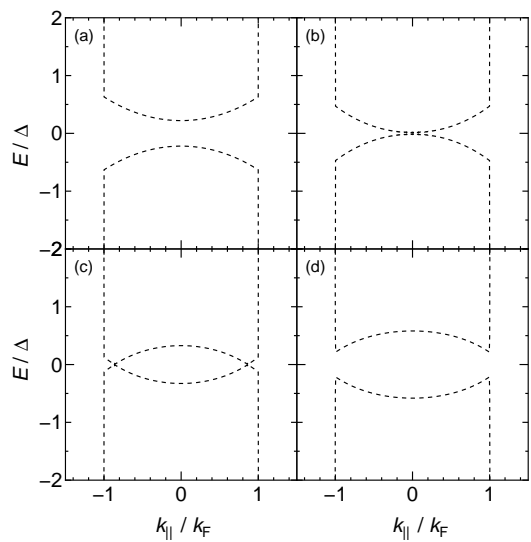}
 \end{center}
 \caption{Energy spectra of STI in $\Delta_{1b}$. Chemical potential $\mu_S$ is set to (a) 0.9 eV, (b)
0.7 eV, (c) 0.5 eV and (d) 0.4 eV. The dotted lines show bulk energy gap. There is no ABS.}
 \label{fig2}
\end{figure}

\begin{figure}[htbp]
 \begin{center}
  \includegraphics[width = 70 mm]{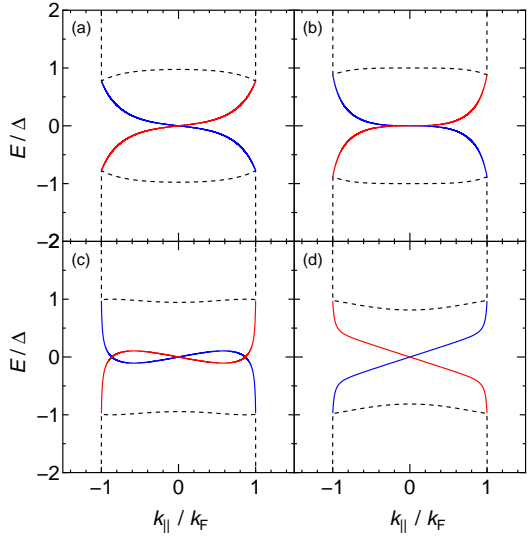}
 \end{center}
 \caption{(Color online) Energy spectra of STI in $\Delta_2$. Chemical potential $\mu_S$ is set to (a)
0.9 eV, (b) 0.7 eV, (c) 0.5 eV and (d) 0.4 eV. The dotted lines show bulk energy gap and
the solid lines show the ABS.}
 \label{fig3}
\end{figure}

\begin{figure}[htbp]
 \begin{center}
  \includegraphics[width = 70 mm]{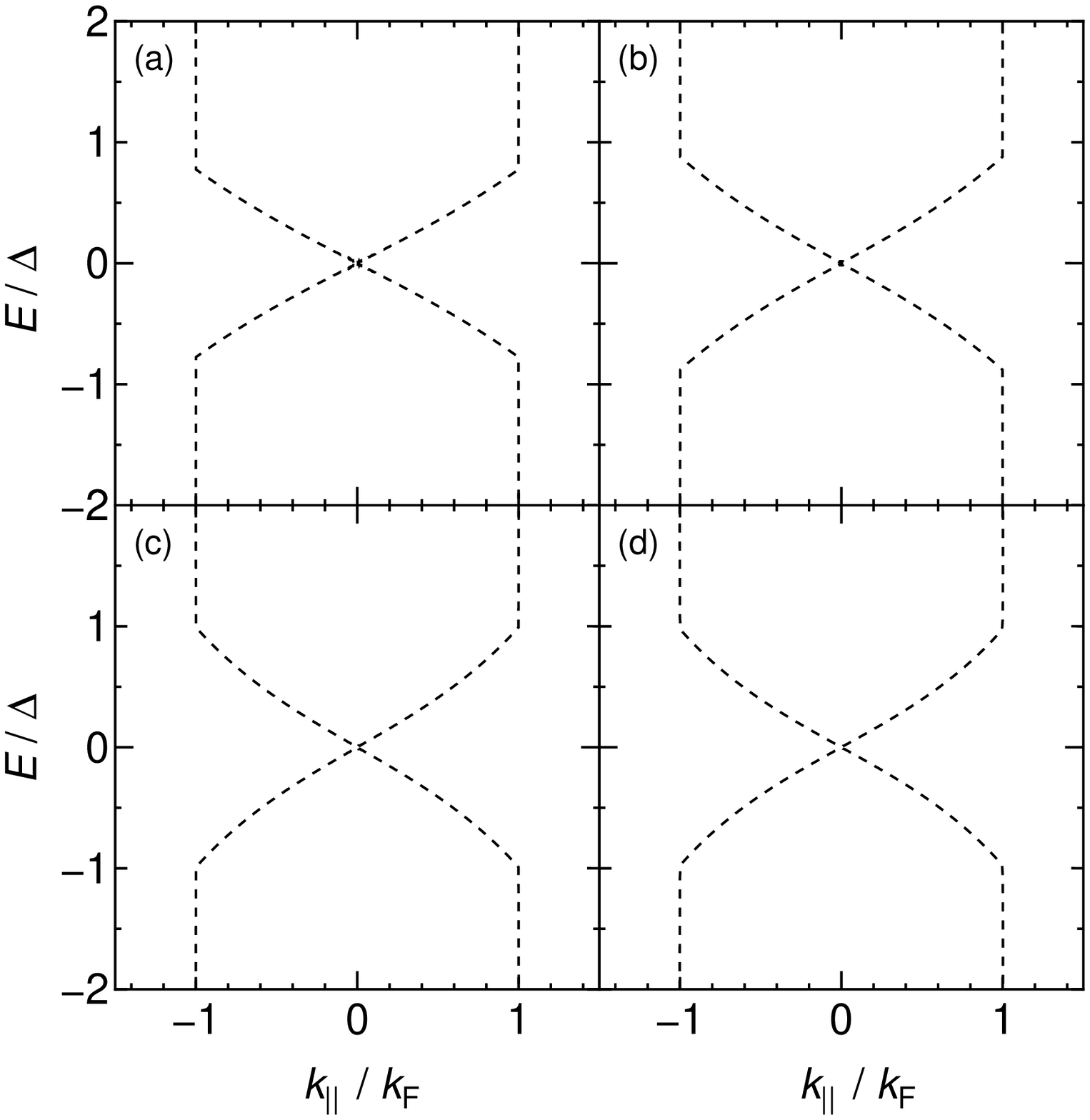}
 \end{center}
 \caption{Energy spectra of STI in $\Delta_3$. Chemical potential $\mu_S$ is set to (a) 0.9 eV, (b)
0.7 eV, (c) 0.5 eV and (d) 0.4 eV. The dotted lines show bulk energy gap. There is no ABS.
The energy gap has point
node at $k_\parallel = 0$.}
 \label{fig4}
\end{figure}

\begin{figure}[htbp]
 \begin{center}
  \includegraphics[width = 70 mm]{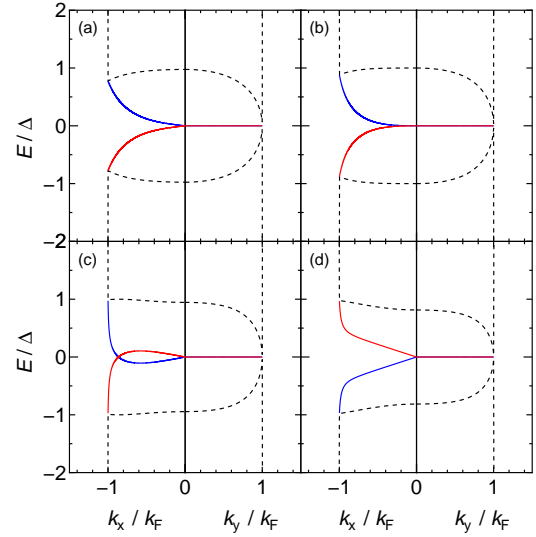}
 \end{center}
 \caption{(Color online) Energy spectra of STI in $\Delta_4$. Chemical potential $\mu_S$ is
set to (a) 0.9 eV, (b) 0.7 eV, (c) 0.5 eV and (d) 0.4 eV. The dotted lines show bulk energy
gap and the solid lines show ABS. In each panel, left (right) side corresponds to the section of $k_y = 0$ $(k_x = 0)$.}
 \label{fig5}
\end{figure}
Now let us see more details of ABSs.
For $\Delta_{1a}$, 
the bulk energy gap is isotropic and there is no ABS.
It is essentially the same with that of conventional 
BCS $s$-wave pairing. 
For the other type of the pair potential,
we show the energy gap of the bulk energy dispersion and ABSs
in Figs. \ref{fig2}-\ref{fig5}. 
Since the energy spectra have rotational symmetry in the $k_{x}$-$k_{y}$ plane, 
we plot bulk energy gap and ABS as a function of $k_\parallel / k_F$ except for $\Delta_{4}$ case.
For $\Delta_{4}$ case, since the spectra does not have this rotational symmetry, 
we plot $E_{k}$ and ABS as a function of $k_{x}/k_{F}$ with $k_{y}=0$ 
in the left side and $k_{y}/k_{F}$ with $k_{x}=0$ in the right side. 
In each figure, the value of the chemical potential $\mu_S$ is chosen to be 
0.9, 0.7, 0.5 and 0.4 eV (which are shown by dots in Fig.1) for 
(a),(b),(c) and (d), respectively.

In the case of $\Delta_{1b}$,
though its irreducible representation is the same as $\Delta_{1a}$,
the bulk energy gap has an anisotropy as seen from Eq. (\ref{eq:1b0}).
Since $m({\bm k}')$ can be zero in the regions II and IV,
line nodes appear in these regions.
Thus, the energy gap closes for $\mu_S=0.5$ as shown in Fig. \ref{fig2}(c).
In other regions, $\Delta_{1b}$ is fully gapped as shown in Figs. \ref{fig2}(a), (b) and (d).
No ABS appears for this gap function as in the case of $\Delta_{1a}$.

In the case of $\Delta_{2}$, the bulk energy dispersion has a fully gapped structures.
ABSs are generated on the surface at $z=0$. 
In Fig. \ref{fig3}, we plot the dispersion of the ABS by solid lines.
As explained in Fig. \ref{fig1},
the line shapes of the dispersion of the ABS changes with the chemical potential.
For $\mu_S=0.9$ and 0.7 eV (Figs. \ref{fig3}(a) and (b)),
the resulting ABS is the standard Majorana cone as shown in the region I.
The group velocity at $k_\parallel=0$ for $\mu_S=0.7$ is closer to zero than that for $\mu_S=0.9$,
since the values of $\mu_S$ and $m_1$ is close to those on the line B in Fig. \ref{fig1}.
Fig. \ref{fig3}(c) demonstrates a caldera-type dispersion in the region II.
At $\mu_S=0.4$ eV, as shown in Fig. \ref{fig3}(d),
the line shape of the dispersion of the ABS is similar to the standard Majorana cone like Fig. \ref{fig3}(a) and (b)
while the sign of the group velocities of the ABS is opposite.


In the case of $\Delta_3$,
the bulk energy gap closes at $k_\parallel=0$ as seen from Figs. \ref{fig4}(a)-(d).
This comes from point nodes at north and south poles on the Fermi surface
In this pair potential,
the parity of the spatial inversion is odd.
On the other hand, the parity of the mirror reflection at $z=0$ is even.
There is no ABS at the surface $z=0$.

The pair potential $\Delta_4$ belongs to the two-dimensional irreducible representation $E_u$,
$\Delta \sigma_y s_x$ and $\Delta \sigma_y s_y$.
In the present paper, we choose $\Delta \sigma_y s_y$.
As seen from Figs. \ref{fig5}(a)-(d),
the bulk energy gap closes along $k_y$-axis where point nodes exist.
In this direction,
the dispersion of the ABS is completely flat with zero energy.
In a manner similar to $\Delta_2$,
the group velocity of the ABS along $k_x$-axis decreases with $\mu_S$.
Then, a ridge-type ABS appears at $\mu_S=0.5$ as shown in Fig. \ref{fig5}(c).
In the other cases, a valley-type ABS is generated.

\subsection{Conductance}

\begin{figure}[htbp]
 \begin{center}
  \includegraphics[width = 70 mm]{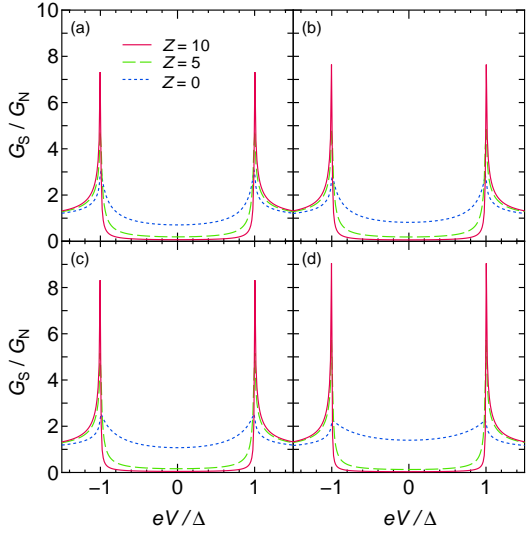}
 \end{center}
 \caption{(Color online) The normalized tunneling conductance in normal metal/STI($\Delta_{1a}$)
junction. Chemical potential is set to 0.2 eV in normal metal and (a) 0.9 eV, (b) 0.7 eV,
(c) 0.5 eV and (d) 0.4 eV in STI. $Z$ is the height of the potential barrier at the interface.}
 \label{fig6}
\end{figure}

\begin{figure}[htbp]
 \begin{center}
  \includegraphics[width = 70 mm]{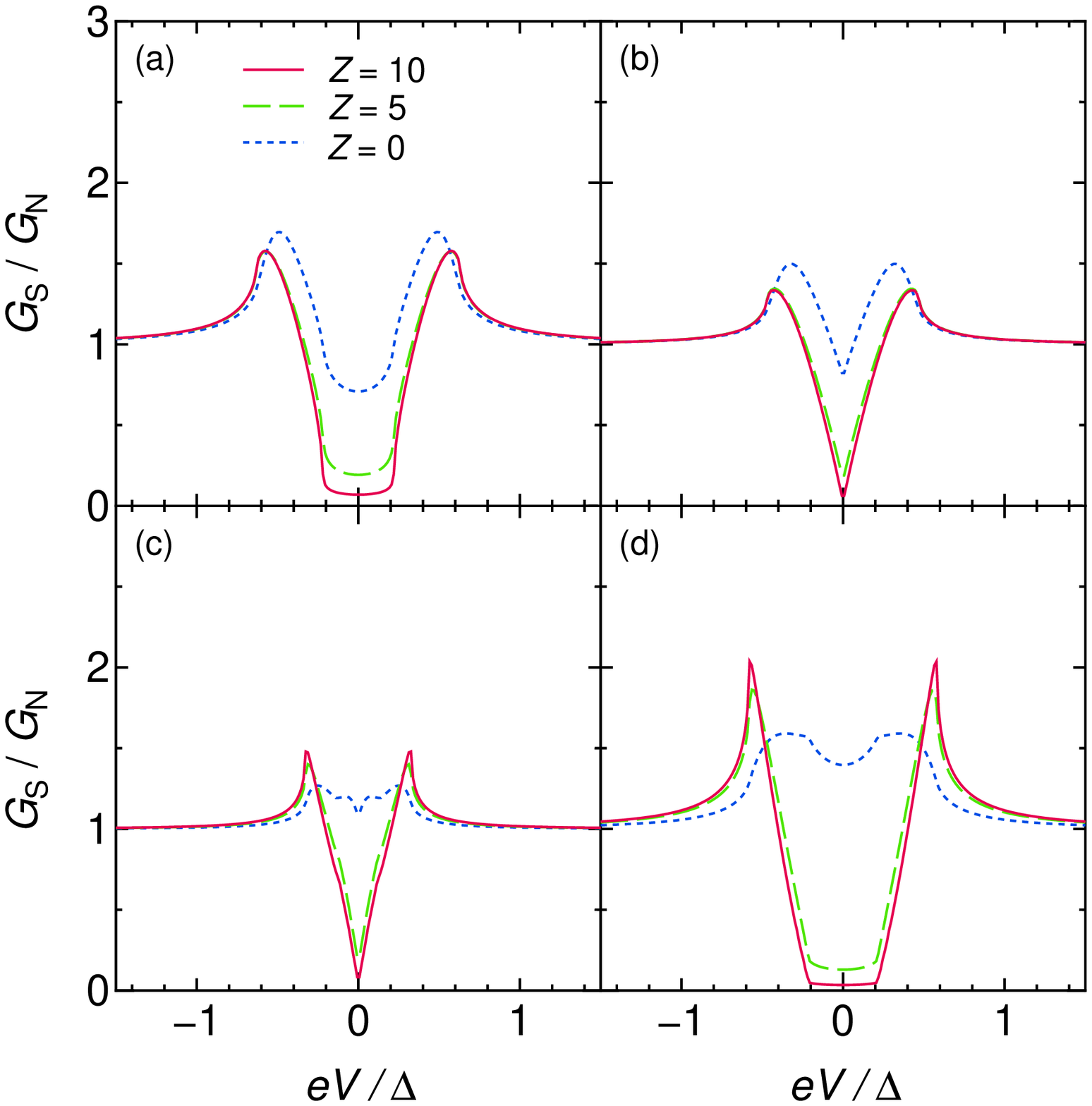}
 \end{center}
 \caption{(Color online) The normalized tunneling conductance in normal metal/STI($\Delta_{1b}$)
junction. Chemical potential is set to 0.2 eV in normal metal and (a) 0.9 eV, (b) 0.7 eV,
(c) 0.5 eV and (d) 0.4 eV in STI. $Z$ is the height of the potential barrier at the interface.}
 \label{fig7}
\end{figure}

\begin{figure}[htbp]
 \begin{center}
  \includegraphics[width = 70 mm]{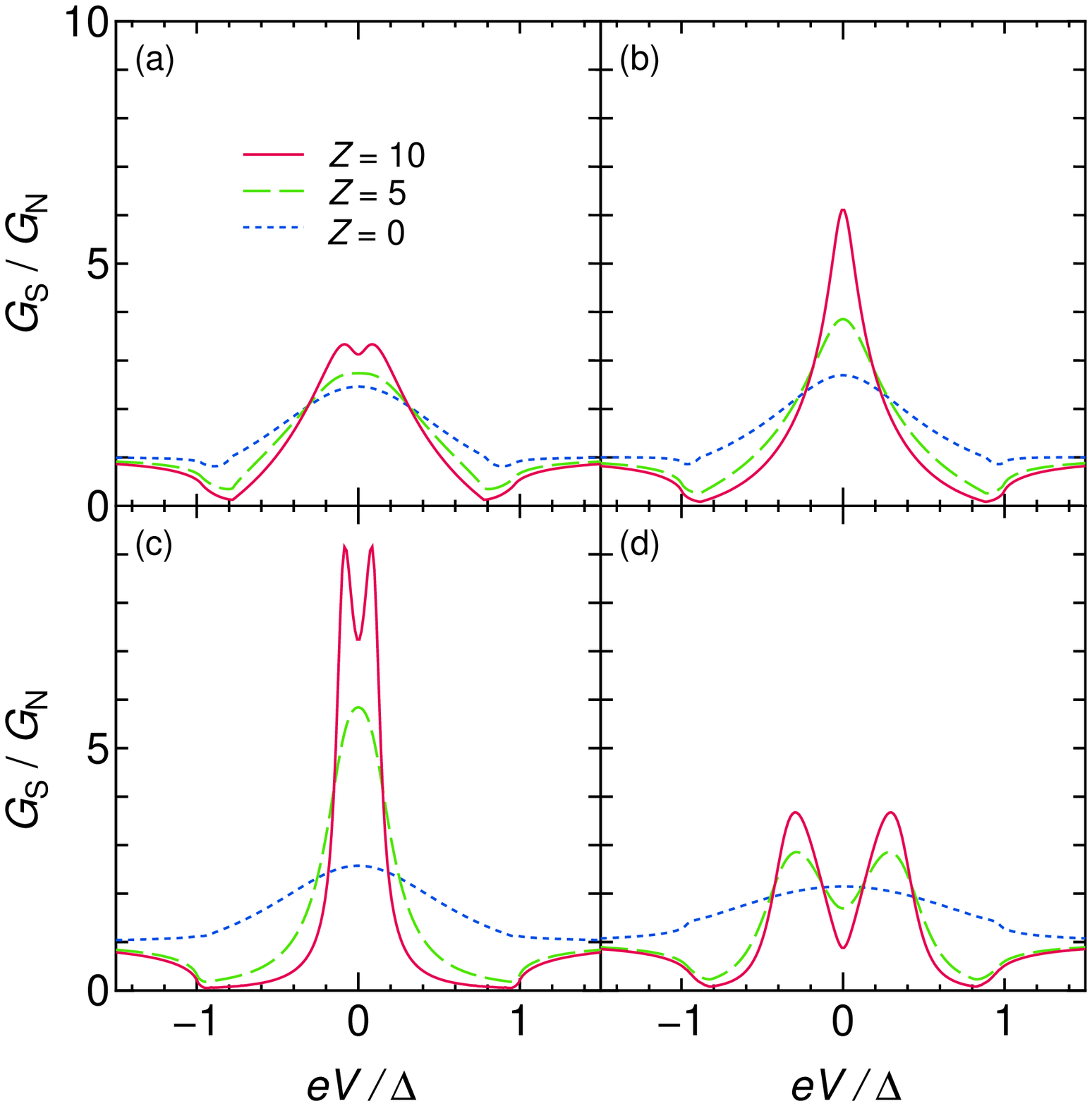}
 \end{center}
 \caption{(Color online) The normalized tunneling conductance in normal metal/STI($\Delta_2$)
junction. Chemical potential is set to 0.2 eV in normal metal and (a) 0.9 eV, (b) 0.7 eV,
(c) 0.5 eV and (d) 0.4 eV in STI. $Z$ is the height of the potential barrier at the interface.}
 \label{fig8}
\end{figure}

\begin{figure}[htbp]
 \begin{center}
  \includegraphics[width = 70 mm]{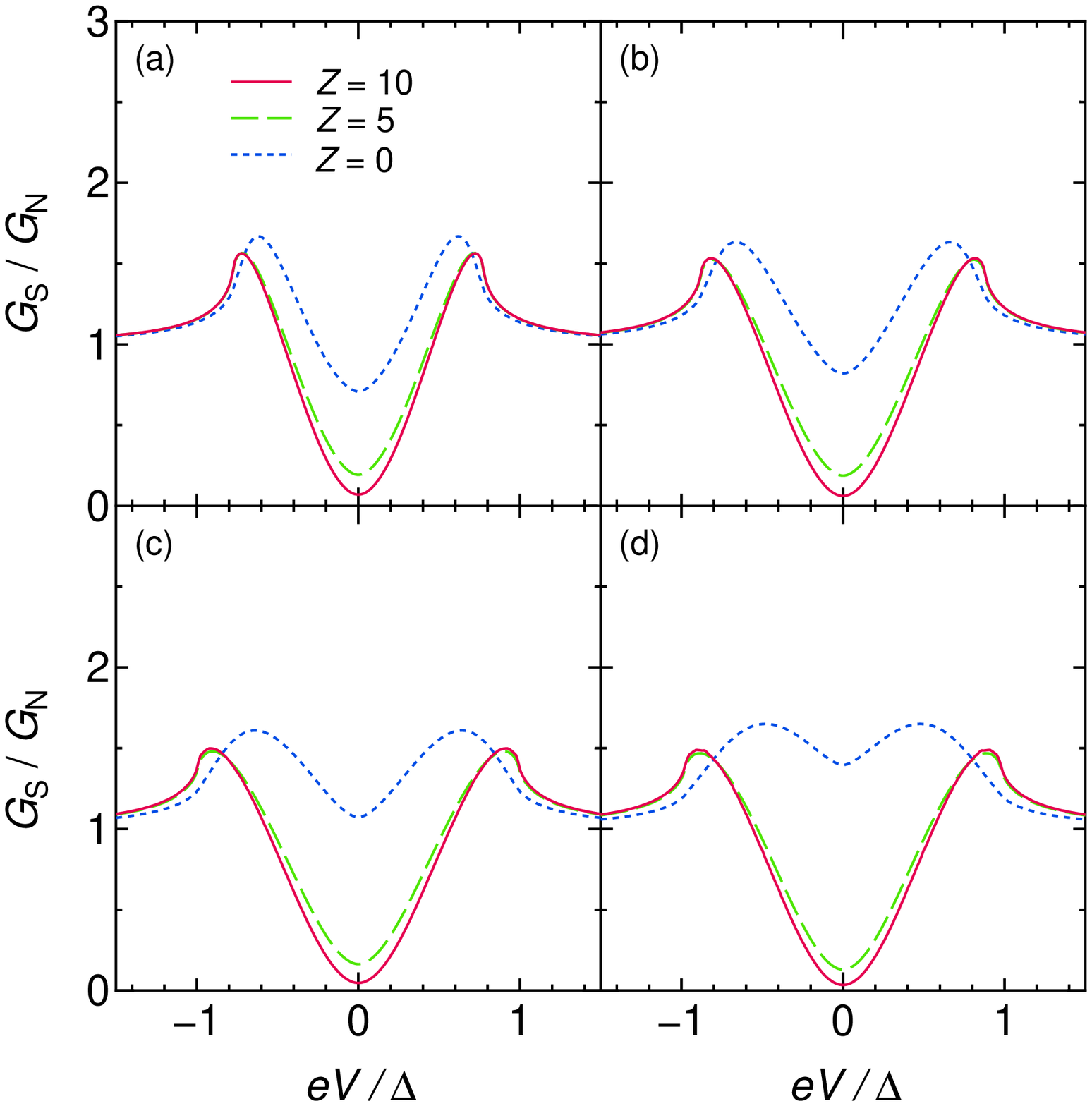}
 \end{center}
 \caption{(Color online) The normalized tunneling conductance in normal metal/STI($\Delta_3$)
junction. Chemical potential is set to 0.2 eV in normal metal and (a) 0.9 eV, (b) 0.7 eV,
(c) 0.5 eV and (d) 0.4 eV in STI. $Z$ is the height of the potential barrier at the interface.}
 \label{fig9}
\end{figure}

\begin{figure}[htbp]
 \begin{center}
  \includegraphics[width = 70 mm]{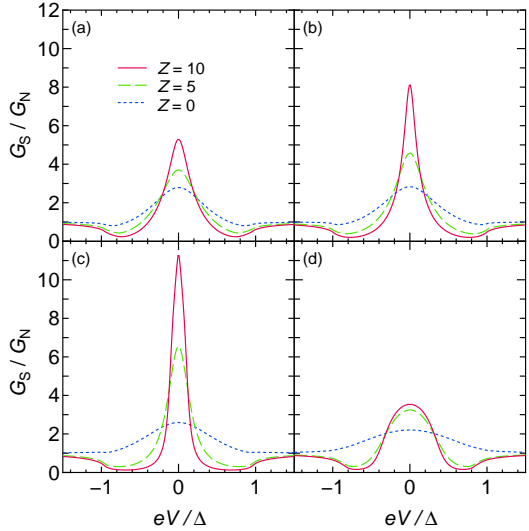}
 \end{center}
 \caption{(Color online) The normalized tunneling conductance in normal metal/STI($\Delta_4$)
junction. Chemical potential is set to 0.2 eV in normal metal and (a) 0.9 eV, (b) 0.7 eV,
(c) 0.5 eV and (d) 0.4 eV in STI. $Z$ is the height of the potential barrier at the interface.}
 \label{fig10}
\end{figure}
In this subsection, we show the bias voltage dependence of tunneling conductance 
for all possible pairings,
 $\Delta_{1a}$, $\Delta_{1b}$, $\Delta_{2}$, $\Delta_{3}$ and $\Delta_{4}$ in
Figs. \ref{fig6}, \ref{fig7}, \ref{fig8}, \ref{fig9} and \ref{fig10}, respectively.
For the magnitude of the barrier potential $Z$,
we choose $Z=0$, 5, and 10 for high, intermediate and low transmissivity, respectively.
$\mu_N$ and $m_N$ are chosen as $\mu_N=0.2$ eV and $\hbar^2/(2m_N)=1$ eV \AA$^2$.
For $\Delta_{1a}$, the obtained conductance rarely
depends on $\mu_S$ qualitatively as shown in Fig. \ref{fig6}. 
For the junction with high transmissivity, a nearly flat nonzero conductance appears around zero voltage.
On the other hand, in the case of low transimissivity, 
the conductance have $U$-shaped structures. 
These features are standard in conventional spin-singlet $s$-wave 
superconductors obtained by BTK theory.
\cite{BTK}

In the case of $\Delta_{1b}$, 
the resulting conductance has a ZBCD
independent of the chemical potential for $Z=0$. 
For $Z=10$, 
the conductance has a $U$-shaped structure for 
(a) $\mu_S=0.9$ eV  and (d) $0.4$ eV. 
On the other hand for (b) $\mu_S= 0.7$ eV and (c) $0.5$ eV, 
we obtain $V$-shaped tunneling conductance
due to the highly anisotropic energy gap.

For $\Delta_{2}$, the tunneling conductance 
shows a simple broad peak around zero voltage for $Z=0$ as shown in Fig. \ref{fig8}.
For $Z=10$, the dispersion of the ABS 
seriously influences the line shape of the tunneling conductance
since the tunneling current flows through the ABSs in the case of low transmissivity.
The conductance shows a ZBCD
except for the case of $\mu_S=0.7$ eV in Fig. \ref{fig8}(b).
A ZBCP appears for $\mu_S=0.7$ eV.
This difference originates from the difference in the dispersion of the ABS:
The dispersion of the ABS shows the standard Majorana cone
like a surface state of BW-phase of superfluid $^{3}$He.
It has been known that,
in the BW-phase,
the tunneling conductance has a ZBCD like a curve for $Z=10$ in Fig. 8(a).\cite{Asano2003,yamakage12}
In the parameter regime near line B in Fig. \ref{fig1}, however,
the group velocity of the ABS around zero energy is almost zero.
Therefore, the surface density of states near the zero energy is enhanced and
the resulting tunneling conductance has a ZBCP. 
On the other hand, if the magnitude of the group velocity is increasing, 
then the surface density of states near zero energy is reduced. 
Then, the surface density of states has a $V$-shaped structure and the 
resulting tunneling conductance can show a ZBCD (see Figs. \ref{fig8}(a), \ref{fig8}(c) and \ref{fig8}(d)).

In the case of $\Delta_{3}$, 
the obtained conductance always have ZBCD as shown in Fig. \ref{fig9}.
Since there is no ABS in the present junction,
conductance for $Z=10$ is proportional to $(eV)^2$ around zero voltage
reflecting the presence of the point nodes on the Fermi surface.

The conductance in the case of $\Delta_4$ shows a ZBCP peak
as shown in Fig. \ref{fig10}.
The existence of the flat zero-energy ABS in the direction of $k\parallel k_y$
induces a ZBCP regardless of the magnitude of the chemical potential.

The calculated results by our analytical formula of conductance well reproduce the preexisting numerical results.
\cite{yamakage12}
This means that the quasi-classical approximation works well in this system.
The reason for this is that the present system has a single Fermi surface,
where simplified calculation is available.
\subsection{Magneto-tunneling conductance}
In this subsection, we study magneto-tunneling spectroscopy 
as an application of this new formula of conductance.
Since both $\Delta_2$ and $\Delta_4$ can have ZBCP,
it is difficult to distinguish between these two pair potentials by simple tunneling spectroscopy.
To resolve this problem,
magneto-tunneling conductance is useful to know the detailed structure of the energy gap and the ABS.

\begin{figure}[htbp]
 \begin{center}
  \includegraphics[width = 70 mm]{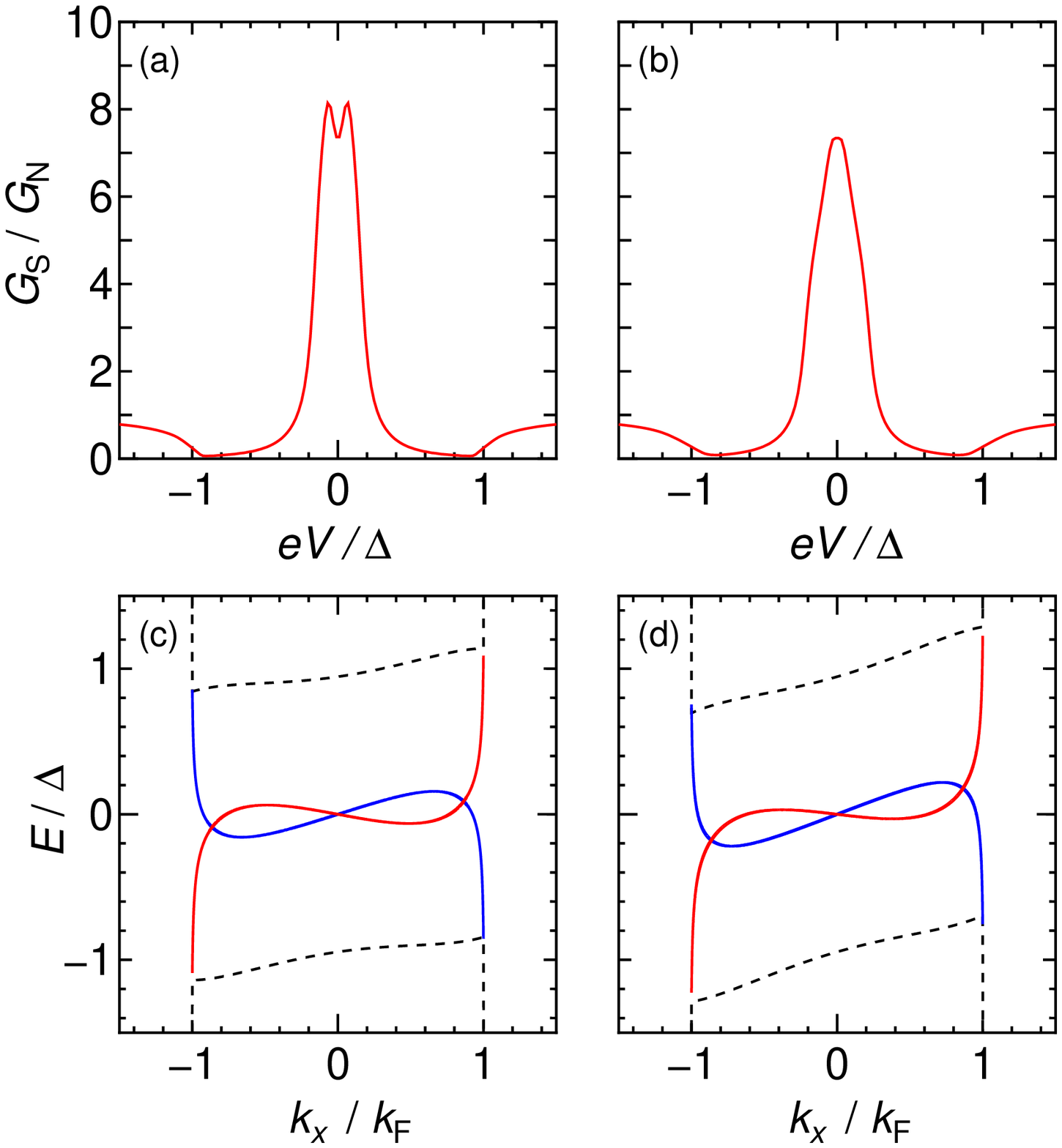}
 \end{center}
 \caption{(Color online) Tunneling conductance and ABS for $\Delta_2$ under an external
magnetic field in the $x$-direction.
We choose $\mu_S=0.5$ eV and $\tilde{H} = e v_\parallel H \lambda / \Delta=$ (a)(c) 0.12 and (b)(d) 0.23.}
 \label{fig11}
\end{figure}

\begin{figure}[htbp]
 \begin{center}
  \includegraphics[width = 70 mm]{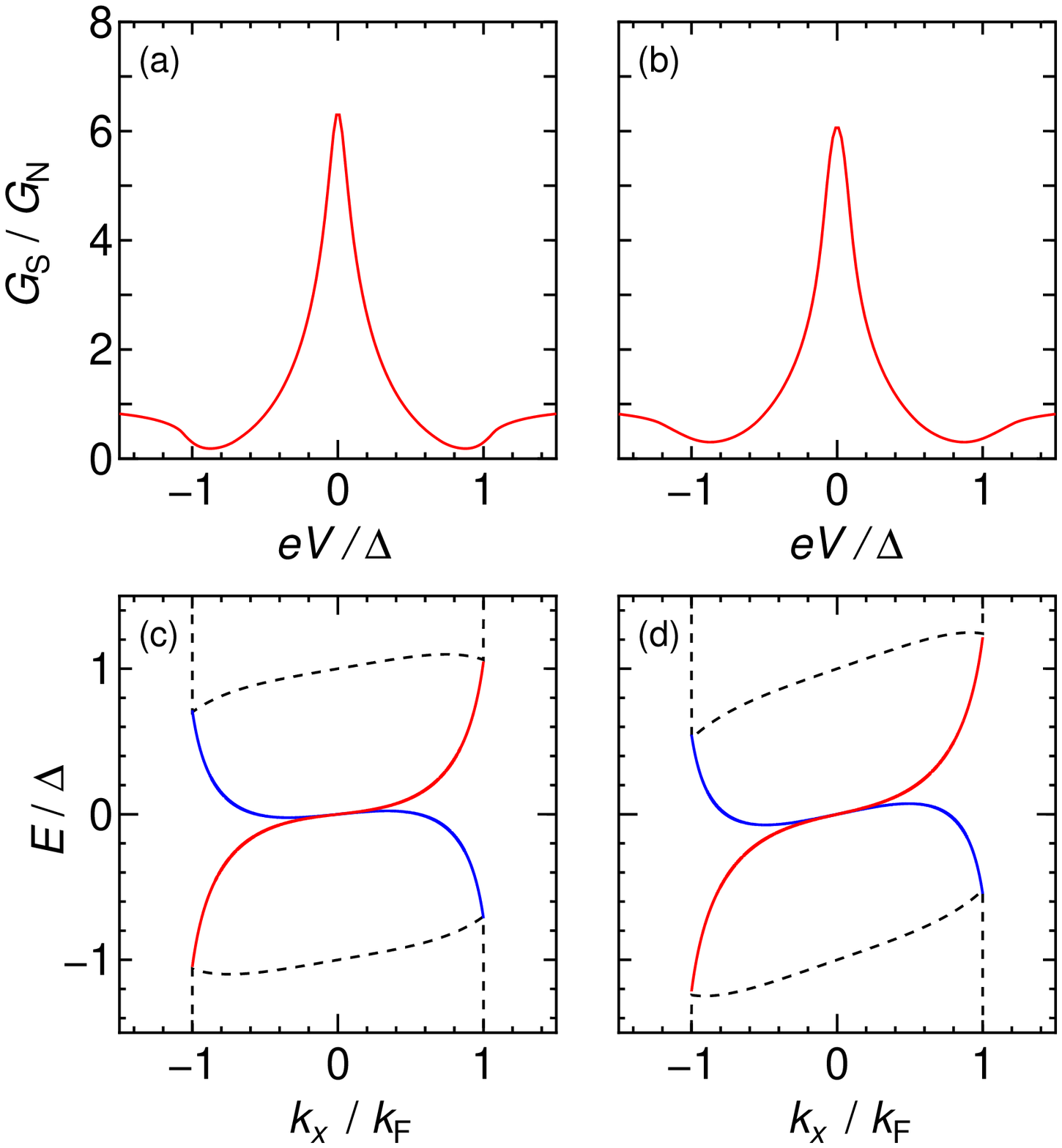}
 \end{center}
 \caption{(Color online) Tunneling conductance and ABS for $\Delta_2$ under an external
magnetic field in the $x$-direction.
We choose $\mu_S=0.7$ eV and $\tilde{H} = ev_\parallel H \lambda / \Delta=$ (a)(c) 0.17 and (b)(d) 0.33.}
 \label{fig12}
\end{figure}

\begin{figure}[htbp]
 \begin{center}
  \includegraphics[width = 70 mm]{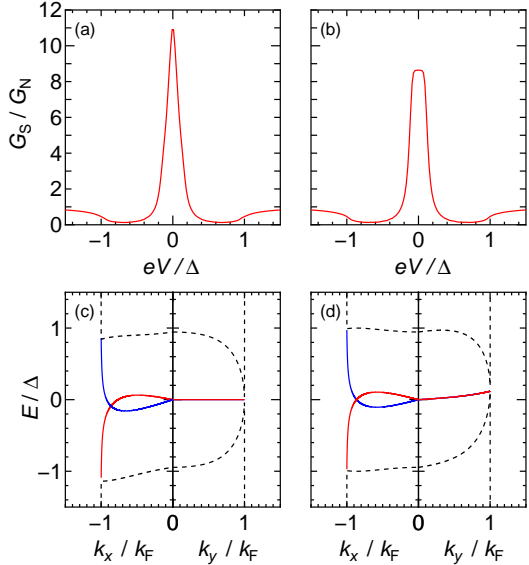}
 \end{center}
 \caption{(Color online) Tunneling conductance and ABS in $\Delta_4$ under an external
magnetic field in (a)(c) the $x$-direction and (b)(d) the $y$-direction.
We choose $\mu_S=0.5$ eV in STI and $\tilde{H} = e v_\parallel H \lambda / \Delta$ is 0.12.}
 \label{fig13}
\end{figure}

\begin{figure}[htbp]
 \begin{center}
  \includegraphics[width = 70 mm]{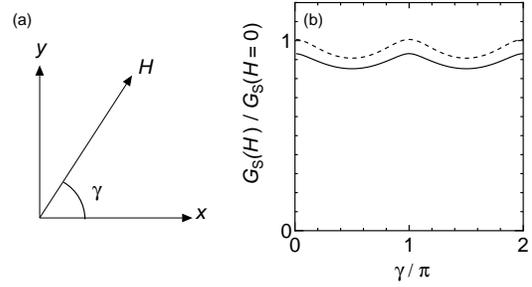}
 \end{center}
 \caption{
(a) Schematic illustration of the direction of external magnetic field.
(b) Zero bias conductance as a function of the rotational angle $\gamma$ in normal metal/STI($\Delta_4$) junction.
Solid line (dotted line) shows the result for $\mu_S=0.5$ eV (0.7 eV).
}
 \label{fig14}
\end{figure}

We show the ABS and the tunneling conductance for $\Delta_2$ with $\mu_S = 0.5 \ \rm eV$
in the presence of in-plane magnetic fields in Fig. \ref{fig11}.
Since the energy dispersion of the quasi-particle is given by
$E({\bm k} - e {\bm A}) =E({\bm k}) - e v_\parallel H \lambda \cos(\phi - \gamma)$,
the magnitude of the Doppler shift is prominent
when the azimuthal angle of the momentum $\phi$ coincides with the direction of the magnetic field $\gamma$.
Thus, the energy dispersion of the ABS is tilted in the direction of the applied magnetic field.
Then, the dispersion of the ABS shown in Fig. \ref{fig3}(c) becomes those in Figs. \ref{fig11}(c) and (d)
for $\tilde{H} = 0.12$ and $0.23$, respectively.
Because the magnitude of the Doppler shift is proportional to that of the applied field,
the value of the group velocity of the one of the edge modes approaches to zero in higher fields,
and thus surface density of states near zero energy increases.
As a result, the ZBCD structure in the conductance is smeared as shown in Figs. \ref{fig11}(a),
and the conductance has a zero bias peak in higher field as shown in Figs. \ref{fig11}(b).
These features have never been seen in Doppler effect in 
high-$T_c$ Cuprate, where the ABS has a flat dispersion 
\cite{Fogel,Tanaka02a,Tanaka02b}. 

Next, we show the case of $\mu_S=0.7$ eV in Fig. \ref{fig12}.
In this case, conductance has a zero bias peak in the absence of the magnetic field as shown in Fig. \ref{fig8}(c)
since the group velocity of the ABS is close to zero.
In the presence of the magnetic field,
group velocities of the two edge channels deviate from zero as shown in \ref{fig12}(c) and (d).
Therefore, the height of the ZBCP decreases with magnetic field.

Next, we show the magneto-tunneling conductance for $\Delta_4$.
This pair potential has an in-plane anisotropy.
Figure \ref{fig13} shows the conductance and the ABS with magnetic field in the $x$-direction ((a), (c))
and the $y$-direction ((b), (d)).
The resulting conductance depends on the direction of the magnetic field.
The height of the ZBCP under magnetic fields in the y-direction (nodal direction) is smaller than that under magnetic fields in the $x$-direction.
This is because the flat ABS in the nodal direction is tilted as shown in Fig. \ref{fig13}(d).
To see the in-plane anisotropy of the magneto-tunneling conductance,
we calculate the magnetic-field angle dependence of the conductance at $eV=0$ in Fig. \ref{fig14}.
It shows minima when the magnetic field is parallel to the nodal direction.\cite{Vekhter}
The angular dependence of the conductance appears only for $\Delta_4$,
since other pair potentials have an in-plane rotational symmetry.
Thus, we can distinguish between $\Delta_2$ and $\Delta_4$.

\section{Summary}
In this paper, we have examined the dispersion of surface ABSs of
STI and the
tunneling conductance in normal metal/STI junctions 
by deriving analytical formula based on the quasiclassical approximation.
Our obtained results are consistent with the previous numerical 
calculation by Yamakage $et$ $al.$ \cite{yamakage12} 
which does not use quasiclassical approximation.
By using the obtained analytical formula of tunneling conductance,
one can easily calculate the tunneling conductance
without any special techniques of numerical calculation.
In addition, we have calculated the tunneling conductance under external magnetic fields
in the $xy$-plane by taking account of the Doppler shift.
As a result, we have shown that the pair potential $\Delta_2$ and $\Delta_4$ can be distinguished
by measuring the field-angle dependence of the zero-bias conductance. 
In this paper, we have studied ballistic normal metal / STI junctions. 
The extension of our conductance formula 
to the diffusive normal metal / STI junction by circuit theory \cite{Proximityd}is interesting since we can expect anomalous proximity effect \cite{Proximityp}
by odd-frequency pairing \cite{TanakaGolubov2007}. 

\section*{Acknowledgements}
This work was supported in part
by Grants-in-Aid for Scientific Research from the Ministry
of Education, Culture, Sports, Science and Technology
of Japan "Topological Quantum Phenomena" (Grant
No. 22103005 and No. 25287085) and the Strategic International Cooperative Program (Joint Research Type) from the Japan Science and Technology Agency.

\bibliographystyle{jpsj}
\bibliography{66521}

\end{document}